\documentclass{aa}  

\usepackage{graphicx}
\usepackage{natbib, twoopt}
\usepackage[varg]{txfonts}
\bibpunct{(}{)}{;}{a}{}{,} 

\begin{document} 

\title{The Role of Stellar Radial Motions in Shaping Galaxy
Surface Brightness Profiles}
\titlerunning{Shaping surface brightness profiles via stellar 
radial motions}

\author{T. Ruiz-Lara, \inst{1,2,3,4} C. G. Few, \inst{5,6} 
E. Florido, \inst{3,4} B. K. Gibson, \inst{5,6}
I. P\'erez, \inst{3,4} \and P. S\'anchez-Bl\'azquez \inst{7}}

\authorrunning{T. Ruiz-Lara et~al.}

\institute{\inst{1} Instituto de Astrof\'isica de Canarias, Calle V\'ia L\'actea s/n, E-38205 La Laguna, Tenerife, Spain \\
\email{tomasruizlara@gmail.com} \\
\inst{2} Departamento de Astrof\'isica, Universidad de La Laguna, E-38200 La Laguna, Tenerife, Spain \\ 
\inst{3} Departamento de F\'isica Te\'orica y del Cosmos, Universidad de Granada, Campus de Fuentenueva, E-18071 Granada, Spain \\
\inst{4} Instituto Carlos I de F\'isica Te\'orica y computacional, Universidad de Granada, E-18071 Granada, Spain \\
\inst{5} E.A. Milne Centre for Astrophysics, University of Hull, Hull HU6 7RX, UK \\
\inst{6} Joint Institute for Nuclear Astrophysics, Center for the Evolution of the Elements (JINA-CEE) \\
\inst{7} Departamento de F\'isica Te\'orica, Universidad Aut\'onoma de Madrid, E-28049 Cantoblanco, Spain \\} 
    
\date{Received ---; accepted ---}

 
\abstract
{}
{The physics driving features such as breaks 
observed in galaxy surface brightness (SB) profiles 
remains contentious. Here, we assess the importance of 
stellar radial motions in shaping their characteristics.}
{We use the simulated Milky Way-mass, cosmological discs, from 
the Ramses Disc Environment Study ({\tt RaDES}) to characterise the 
radial redistribution of stars in galaxies displaying type I (pure 
exponentials), II (downbending), and III (upbending) SB profiles. We 
compare radial profiles of the mass fractions and the velocity dispersions of different 
sub-populations of stars according to their birth and current 
locations.}
{Radial redistribution of stars is important in all galaxies 
regardless of their light profiles. Type II breaks seem to be a consequence of the 
combined effects of outward-moving and accreted stars. The 
former produces shallower inner profiles (lack of stars in the inner 
disc) and accumulate material around the break radius and beyond,
strengthening the break; the latter can weaken or even convert
the break into a pure exponential. Further accretion from satellites 
can concentrate material in the outermost parts, leading to type III 
breaks that can coexist with type II breaks, but situated further 
out. Type III galaxies would be the result of 
an important radial redistribution of material throughout the entire disc,
as well as a concentration of accreted material in the outskirts.
In addition, type III 
galaxies display the most efficient radial redistribution and the 
largest number of accreted stars, followed by type I and II systems, 
suggesting that type I galaxies may be an intermediate case 
between types II and III. In general, the velocity dispersion profiles of all galaxies 
tend to flatten or even increase around the locations where the breaks are found. The age and 
metallicity profiles are also affected, exhibiting different inner gradients depending on 
their SB profile, being steeper in the case of type II systems (as 
found observationally). The steep type II profiles 
might be inherent to their formation rather than acquired via radial 
redistribution.}
{}


\maketitle

\section{Introduction}
The observed properties of spiral galaxies are the outcome of complex, 
non-linear, and inter-related, formation and evolution processes. 
According to the current paradigm, after the assembly of large galaxies 
via a number of high-redshift minor mergers, a period of secular 
evolution follows. In parallel with the latter, non-axisymmetric 
structures such as bars or spiral arms may be formed 
\citep[e.g.][]{1978MNRAS.183..341W, 2003ApJ...591..499A} with the 
ability to redistribute material \citep[][]{2002MNRAS.336..785S, 
2008ApJ...675L..65R, 2008ApJ...684L..79R, 2010ApJ...722..112M} across 
the entire galaxy. Simultaneous to this secular evolution, satellite 
accretion may continue to influence the characteristics of the host 
\citep[][]{2007ApJ...670..269Y, 2012MNRAS.420..913B, 
2016A&A...586A.112R}, particularly those found in the outermost regions.

A growing number of theoretical and observational works are focused on 
analysing the characteristics of these outer parts of spiral galaxies 
\citep[e.g.][]{2008ApJ...675L..65R, 2008ApJ...684L..79R, 
2009MNRAS.398..591S, 2012A&A...548A.126M, 2008ApJ...683L.103B, 
2012ApJ...752...97Y, 2016MNRAS.456L..35R}. The lower surface densities 
(and, hence, gravitational effects) and the longer dynamical times of 
stars populating those regions make the study of the outskirts of disc 
galaxies a unique place to test current galaxy formation and evolution 
models. One of the defining metrics of these outer regions is the 
presence, or lack thereof, of deviations from a pure exponential surface 
brightness (SB) profile \citep[][]{1940BHarO.914....9P, 
1970ApJ...160..811F, 2005ApJ...629..239B, 1987A&A...173...59V, 
2002A&A...392..807P, 2004A&A...427L..17P, 2005ApJ...630L..17T, 
2005ApJ...626L..81E, 2006A&A...455..467F, 2012MNRAS.427.1102M}. While 
some galaxies possess an essentially pure exponential SB profile into 
the outer parts (type I, continuation of the inner behaviour), others 
show a deficiency (type II) or excess of light (type III) that can be 
characterised by two exponentials \citep[][]{2006A&A...454..759P}. The 
drivers behind these different profiles remain unclear, although changes 
in the ages of stellar populations, the effect of radial migration, and 
satellite accretion have all been proposed 
\citep[][]{2007ApJ...670..269Y, 2008ApJ...675L..65R, 
2009MNRAS.398..591S, 2009ApJ...705L.133M, 2012A&A...548A.126M}.

In the last decade a clear link between an outer upturn (``U-shape'') in 
the age profiles and type II galaxies was found via observations and 
theory, with the age upturn causing the lack of light in the outer 
regions \citep[e.g.][]{2008ApJ...683L.103B, 2008ApJ...675L..65R, 
2009MNRAS.398..591S, 2009ApJ...705L.133M, 2012ApJ...752...97Y}. 
Conversely, in \citet[][]{2016MNRAS.456L..35R}, analysing spectroscopic 
information from the CALIFA\footnote{\url{http://califa.caha.es/}} 
\citep[Calar Alto Legacy Integral Field Area,][]{2012A&A...538A...8S} 
survey, we found that ``U-shape'' age profiles are found for both type I 
and II galaxies \citep[similar to that found by][ from photometric 
data]{2012ApJ...758...41R}. As a consequence, we suggested that the 
mechanisms causing the light distributions and the stellar age profiles 
might not be coupled. In the same vein, \citet[][]{2016A&A...586A.112R} 
showed that all galaxies in the {\tt RaDES} \citep[Ramses Disc 
Environment Study,][]{2012A&A...547A..63F} simulation suite show 
``U-shape'' stellar age profiles, regardless of their light 
distribution. Other observational studies have also found the presence 
of old stellar populations in the outer parts of nearby spiral galaxies 
\citep[e.g.][]{2010ApJ...712..858G, 2012MNRAS.420.2625B, 
2015MNRAS.446.2789B}. All these works seem to indicate that a change in 
the stellar population age is not necessarily the main agent shaping the 
observed light profiles.

Other studies have concentrated on the role of satellite bombardment in 
shaping light profiles \citep[e.g.][]{2007ApJ...670..269Y, 
2009MNRAS.397.1599Q, 2012MNRAS.420..913B}. For instance, 
\citet[][]{2007ApJ...670..269Y} show that minor mergers can cause type 
III SB profiles by concentrating mass in the inner regions or expanding 
the outer discs. However, most of the effort to understand the 
occurrence of the different SB profiles using simulations has been 
focused on analysing the effect of stellar radial motions/redistribution 
\citep[][]{2008ApJ...675L..65R, 2009MNRAS.398..591S, 
2009ApJ...705L.133M, 2012A&A...548A.126M, 2016MNRAS.460L..94G, 2016ApJ...830L..40S}.

\citet[][]{2009MNRAS.398..591S} found that downbending light profiles 
(type II) arise (together with ``U-shape'' age profiles) from the 
combined effect of i) an abrupt change in the radial star formation 
profile due to a change in the gas volume density profile linked to a 
warped disc (causing the truncation in the light profile) and ii) radial 
migration of stars formed in the inner parts towards positions located 
beyond the break radius. The authors claim that it is the first effect 
(flaring of the disc) which drives the ``U-shape'' age and downbending 
light profiles, while the second effect modifies the final shape of such 
profiles. However, they also speculate that different SB profiles might 
arise as a consequence of differences in the efficiency of both 
processes. In this way, if the distribution of gas volume density 
changes smoothly with galactocentric distance and the outwards radial 
redistribution of material is more efficient than in a type II galaxy, 
one might obtain a type III profile (upbending light profile). The same 
conditions might be applicable to pure exponential discs if the radial 
redistribution of material is less efficient than in type III systems. 
Indeed, U-shaped age profiles can be a natural outcome in analytical 
models of pure exponentials with radially varying gas infall 
prescriptions, even in the complete absence of radial stellar (or gas) 
motions \citep[][]{BuenosAires}.  In a similar line of reasoning, 
\citet[][]{2017A&A...604A...4R} recently found different stellar age and 
metallicity inner gradients for galaxies displaying type I, II, and III 
SB profiles. They interpret those results as the outcome of a gradual 
increase in the radial redistribution efficiency from type II to type I 
and III galaxies.\footnote{An interpretation, admittedly, lacking a 
theoretical framework.}

Recently, several studies have tried to shed further light onto the 
effect of radial redistribution of material in shaping SB profiles. 
\citet[][]{2015MNRAS.448L..99H, 2017MNRAS.470.4941H} investigated the 
role of the halo spin parameter ($\lambda$) in shaping the outer SB 
profiles by analysing a set of controlled simulations of isolated 
galaxies. They found a clear transition from type III systems displaying 
low spin parameters to type II galaxies showing higher values with type 
I discs having intermediate values ($\lambda$ $\sim$ 0.035). In 
particular, they suggested that orbital resonances with a strong central bar, 
coupled with the low initial halo spin, 
can produce stellar migration which leads to upbending 
SB profiles. According to this work, stars populating these outer regions 
present high radial velocity dispersions and a lower degree of rotation than expected 
in disc-like systems. \citet[][]{2012A&A...548A.126M} found that after 2 Gyr of smooth, 
in-plane gas accretion, galaxies displaying typical type II profiles acquire light upbendings 
in the outermost regions. This results in galaxies displaying a combination of an inner type II 
and an outer type III profiles with the latter being populated by stars that also present high 
velocity dispersions. \citet[][]{2016A&A...591L...7B} revisit the idea that 
flares in galaxies can give rise to discs following a type II SB 
distribution. In \citet[][]{2016ApJ...830..115E}, the authors present 
numerical experiments on the effect of stochastic scattering of random 
particles shaping single exponentials discs. They also report that type 
II and III profiles could be found if a difference in the scattering 
bias for the inner and outer regions exists.

In this work, we make use of the {\tt RaDES} suite of simulated galaxies 
to investigate two of the processes that can drive the different types 
of SB profiles: the effect of radial motions of stars and satellite 
accretion. In \S\ref{simulations}, we outline the simulations employed 
and characterise the associated galaxies.  The main results and 
discussion are given in \S\ref{result1} and \S\ref{result2}; our 
conclusions are presented in \S\ref{conclusions}.

\section{Simulations and sample of galaxies}
\label{simulations}

The {\tt RaDES} galaxies are simulated using the adaptive mesh 
refinement code {\textsc{ramses}} \citep[][]{2002A&A...385..337T} 
tracking dark matter, stars and gas, on cosmological scales. The 
hydrodynamical evolution of gas uses a refining grid such that the 
resolution of the grid evolves to follow overdensities, reaching a peak 
resolution of 436~pc (16 levels of refinement). In order to prevent 
numerical collapse, a polytropic equation of state is used for dense 
gas. If gas is more dense than 0.1~cm$^{-3}$ star formation occurs at a 
rate given by $\dot{\rho} = -\rho/t_{\star}$, where $t_{\star} = 
t_0(\rho/\rho_0$)$^{-1/2}$ with $t_0=8$~Gyr. Stellar feedback is delayed 
to occur $10^7$ years after star formation whereupon it distributes 
kinetic energy, mass and metals to the gas within a two grid cell radius 
sphere. The mass fraction of stellar particles that explode as 
supernovae (SNe) is 10\%, with each SN providing 10$^{51}$ erg of 
energy, and the 10\% of non-metals (H and He) are converted to metals. 
The cosmological parameters employed in generating these realisations 
were: H$_0$=70~km s$^{-1}$Mpc$^{-1}$, $\Omega_{\mathrm{m}}$=0.28, 
$\Omega_{\mathrm{\Lambda}}$=0.72, $\Omega_{\mathrm{b}}$=0.045, and 
$\sigma_8$=0.8.\footnote{Being H$_0$ the Hubble constant, 
$\Omega_{\mathrm{m}}$ the fraction of total matter, 
$\Omega_{\mathrm{\Lambda}}$ the fraction of the dark energy, and 
$\sigma_8$ the strength of the primordial density fluctuations} Two 
different box volumes sizes of 20 h$^{-1}$ Mpc and 24 h$^{-1}$ Mpc were 
used. The mass resolution of dark matter particles was either 
5.5$\times$10$^{6}$~M$_\odot$ or 9.5$\times$10$^{6}$~M$_\odot$, 
respectively, for each of aforementioned two volumes. Further details of 
the halo selection process and the simulation parameters may be found in 
\citet[][]{2012A&A...547A..63F}.

The \tt RaDES \rm suite are a powerful tool in assessing the effect of 
radial motions (including satellite accretion) in shaping SB profiles. 
In \citet[][]{2012A&A...547A..63F}, we showed that these Milky-Way-mass, 
disc-dominated galaxies, possess characteristics resembling those of 
observed systems, such as metallicity gradients, matter content (total, 
dark, stellar, baryonic, and gaseous mass), and rotation curves 
\citep[see also][]{2012A&A...540A..56P}. Unlike other comparable 
simulated cosmological samples, the {\tt RaDES} galaxies are somewhat 
unique in possessing the full range of SB profiles 
\citep[I, II, and III;][]{2016A&A...586A.112R}.

\subsection{Characterisation of the {\tt RaDES} galaxies light distribution}
\label{SB_prof_char}

The light profiles that we analyse in this work have been previously 
shown in \citet[][]{2016A&A...586A.112R}. The radial profiles were 
computed by azimuthally-averaging the light distribution from the mock 
images presented in~\citep[][]{2012A&A...547A..63F}. These images were 
produced using the {\tt SUNRISE} \citep[][]{2006MNRAS.372....2J} code 
mimicking the SDSS bandpasses. {\tt SUNRISE} uses the stellar and 
gaseous distributions, as well as Spectral Energy Distributions (SEDs) 
for each composite stellar particle, drawn from the {\tt Starburst99} 
stellar population models \citep[][]{1999ApJS..123....3L}, in order to 
generate the bandpass-dependent mock images. The SB profiles computed in 
this way were then fitted with the function presented 
in~\citep[][equations 5 and 6]{2008AJ....135...20E} with a broken 
exponential profile implemented. This allows us to characterise the 
light distribution in these simulated galaxies in a similar fashion as 
usually done with observed photometric data (see 
Table~\ref{tab:galaxies}). In the following analysis we will concentrate 
on the results using the SDSS $r$-band light profiles, although 
identical results are found with the other two filters analysed in 
\citet[][]{2016A&A...586A.112R} (SDSS $g$ and $i$-bands).

\begin{table}
\centering
\begin{tabular}{lrrrrr} 
\hline 
Galaxy & SB type & $\rm h_{\rm in}$ & $\rm h_\mathrm{out}$ & $\rm R_\mathrm{break}$ & $\beta$ \\
 & & (kpc) & (kpc) & (kpc) & (kpc) \\\hline 
Apollo    &  II &  2.34 & 1.39 & 4.96 &  0.95 \\
Artemis   & III &  0.79 & 5.87 & 5.30 & -5.08  \\
Atlas     &  II &  4.39 & 2.47 & 7.43 &  1.92  \\
Ben       &  II &  5.28 & 3.36 & 12.18 & 1.93 \\
Castor    &  II &  5.70 & 0.96 & 5.35 &  4.74 \\
Daphne    & III &  1.26 & 3.84 & 10.08 & -2.58 \\
Eos       & III &  2.95 & 8.55 & 18.32 & -5.60 \\
Helios    & III &  1.93 & 7.76 & 11.49 & -5.83 \\
Hyperion  &  II &  4.31 & 2.77 & 14.97 & 1.54 \\
Krios     & III &  2.62 & 8.45 & 16.11 & -5.83 \\
Leia      & III &  3.81 & 7.48 & 21.21 & -3.67 \\
Leto      & III &  0.99 & 4.07 & 6.59 &  -3.08 \\
Luke      &   I &  5.78 &  -   &  -    &  0.00   \\
Oceanus   &  II &  8.08 & 4.18 & 23.17 & 3.90 \\
Pollux    & III &  1.25 & 4.03 & 8.61 &  -2.78 \\
Selene    &  II &  5.68 & 2.01 & 11.98 & 3.66 \\
Tethys    &  II &  4.43 & 2.12 & 9.82 &  2.31 \\
Tyndareus & III &  1.33 & 3.39 & 7.12 & -2.06  \\
Zeus      &   I &  0.92 &  -   &  -    & 0.00  \\
\hline
\end{tabular}
\caption{Main disc properties for the {\tt RaDES} galaxies from the 
analysis of their r-SDSS band light distribution. First column: Galaxy 
name. Second column: Surface brightness type according to the 
\citet[][]{2006A&A...454..759P} classification. Third column: Inner 
disc scale-length in kpc. Fourth column: Outer disc scale-length in 
kpc. Fifth column: Break radius in kpc. Sixth column: strength of the 
break ($\beta$) defined as h$_{\rm in}$~-~h$_{\rm out}$ in kpc.}
\label{tab:galaxies}
\end{table}

A visual morphological classification from the mock images of these 
simulated galaxies suggests that the {\tt RaDES} systems are mainly 
late-type disc galaxies. Thus, we have decided to compare these results 
with those found in \citet[][]{2006A&A...454..759P} analysing 
1-dimensional SB profiles of a sample of 98 late-type spiral galaxies in 
a similar way than in this work to ensure that these simulated galaxies 
display realistic light distributions. Table~\ref{tab:galaxies} shows 
the values of the main parameters describing the light profiles for all 
the {\tt RaDES} galaxies. The average values of the inner disc 
scale-length (h$_{\rm in}$) that we obtain for the {\tt RaDES} galaxies 
for type I, II, and III galaxies are $\sim$ 3.3~$\pm$~1.2, 
5.0~$\pm$~0.8, and 1.9~$\pm$~0.5 kpc, respectively. These values are in 
agreement (within errors) with those found in 
\citet[][]{2006A&A...454..759P} (2.8~$\pm$~0.8, 3.8~$\pm$~1.2, and 
1.9~$\pm$~0.6 kpc for types I, II, and III disc galaxies). Similar 
claims can be outlined regarding the position of the break. The {\tt 
RaDES} type II and III galaxies present values of R$_{\rm 
break}$/h$_{\rm in}$ of 2.2~$\pm$~0.4 and 6.4~$\pm$~0.4, respectively 
while the values from the \citet[][]{2006A&A...454..759P} work are 
1.9~$\pm$~0.6 and 4.9~$\pm$~0.6. Finally, we also find a good agreement 
if we consider the outer disc scale-length (h$_{\rm out}$). The values 
of h$_{\rm out}$ for the {\tt RaDES} type II and III galaxies are 
2.4~$\pm$~0.5 and 5.9~$\pm$~1.0 kpc, respectively, while for the 
\citet[][]{2006A&A...454..759P} sample those values are 2.1~$\pm$~0.9 
and 3.6~$\pm$~1.2 kpc, respectively. Greater differences are found for 
the h$_{\rm out}$ and R$_{\rm break}$/h$_{\rm in}$ values for the type 
III galaxies. These discrepancies might arise as a consequence of the 
well known `angular momentum problem' 
\citep[e.g.][]{2002NewA....7..155S} causing the over-production of the 
spheroid component due to enhanced star formation at early epochs in 
cosmological simulations. Thus, we can claim that, although there are 
some small discrepancies with the typical observed SB profiles 
(especially for type III galaxies), the light distributions in the {\tt 
RaDES} set of galaxies are quite realistic and consistent with 
observations.

In order to establish a continuous parameter to characterise these SB 
profiles, from downbending to upbending (with pure exponentials as 
intermediate cases), we define the strength of the break ($\beta$) as 
h$_{\rm in}$~-~h$_{\rm out}$. These $\beta$ values are shown in 
Table~\ref{tab:galaxies} as well. We must highlight the continuity of 
this parameter, with the {\tt RaDES} set of galaxies being comprised by 
galaxies presenting weak and strong type II and III breaks as well as 
pure exponentials. We should note that all {\tt RaDES} type II galaxies 
also show an outer type III break whose origin will be also analysed in 
the next section \citep[see ][]{2016A&A...586A.112R}. These type-II+type-III 
combined profiles have been also found in \citet[][]{2012A&A...548A.126M} 
as a consequence of in-plane gas accretion. Despite the 
existence of this secondary break, we will focus our SB characterisation 
on the $\beta$ parameter of the break located closer to the centre (with 
$\beta$~=~0 for type I galaxies, negative for type III galaxies, and 
positive for type II galaxies).

In the following we will name each analysed system according to the name 
given in \citet[][]{2012A&A...547A..63F}. For further information in 
each particular galaxy we encourage the reader to check 
\citet[][]{2012A&A...547A..63F}, \citet[][]{2016A&A...586A.112R}, and 
Table~\ref{tab:galaxies}.

\section{Role of radial redistribution in shaping SB profiles}
\label{result1}

\begin{figure*}
\centering\includegraphics[width=0.95\textwidth]{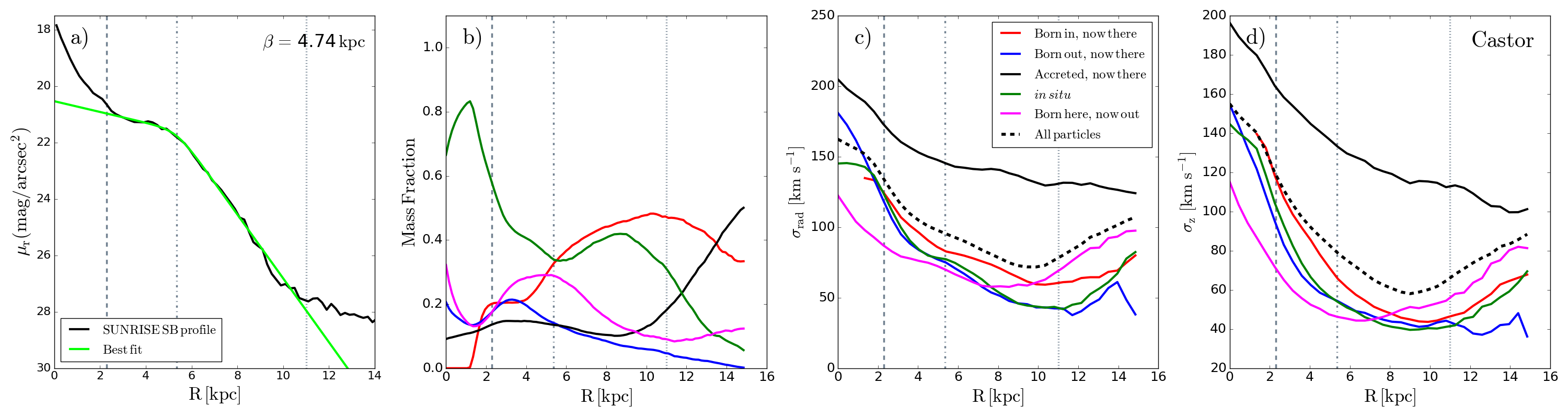} \\
\includegraphics[width=0.95\textwidth]{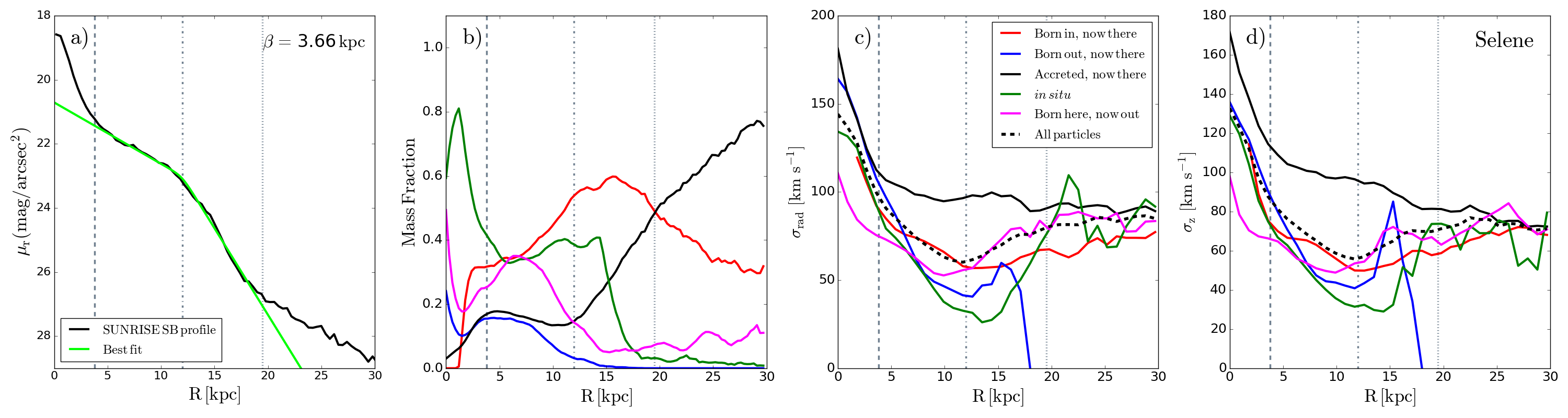} \\
\includegraphics[width=0.95\textwidth]{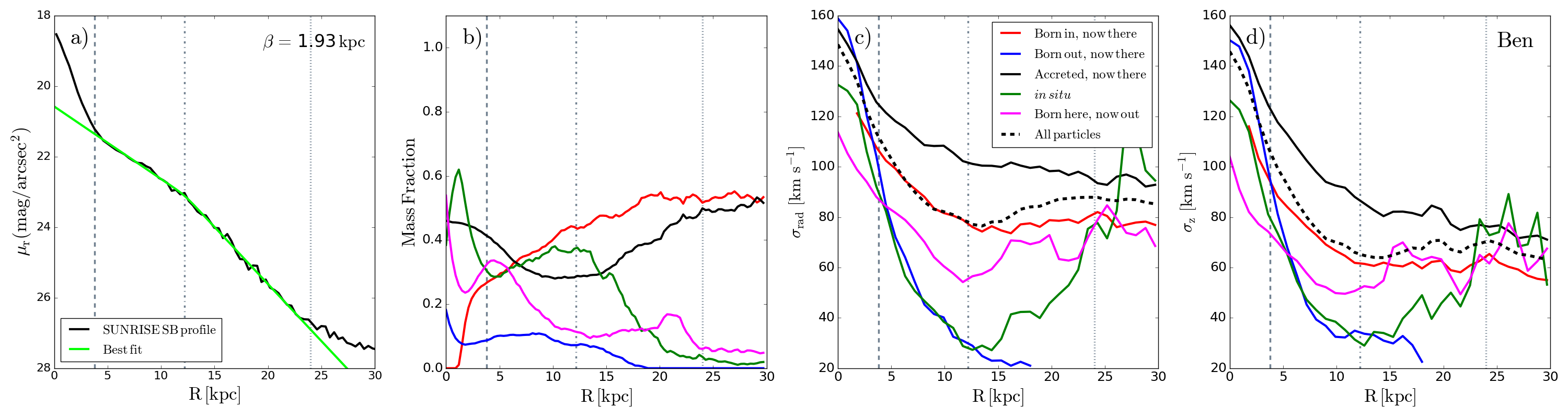} \\
\caption{Characterisation of the light, radial redistribution and velocity 
dispersion profiles of the {\tt RaDES} galaxies. {\it a}) Surface Brightness profiles in SDSS $r$ band 
for 3 type II galaxies with different break strengths (Castor, Selene, 
and Ben). The black solid line represents the simulated 
({\tt SUNRISE}) SB profile while the light-green solid line is the 
best fit of the disc. Afterwards, from left to right: {\it b}) radial redistribution of the stellar mass fraction; 
{\it c}) radial velocity dispersion profiles; 
and {\it d}) vertical velocity dispersion profiles for different subpopulations of stars. The 
different stellar subpopulations analysed include, for these three galaxies, stars not experiencing radial motions (green), 
those born at lower galactocentric distances and now found there (red), those 
born at higher galactocentric distances and now there (blue), those born at 
that radius and currently located at a larger radius (magenta), 
and those coming from satellites (black). We have also included the radial and vertical velocity dispersion profiles 
considering all star particles in the simulation (black dashed lines). $\beta$, defined as h$_{\rm in}$ - h$_{\rm out}$, is 
indicated as an inset. The vertical lines in all panels are located at the end of 
the bulge-dominated region ($\mu_{model}$ - $\mu_{\rm observed}$ $<$ 
0.1 mag arcsec$^{-1}$, gray dashed line), the type II break (gray 
dotted-dashed line), and the type III break (gray dotted line). See 
text for further details.}
\label{fractions_1}
\end{figure*}

\begin{figure*}
\centering
\includegraphics[width=0.95\textwidth]{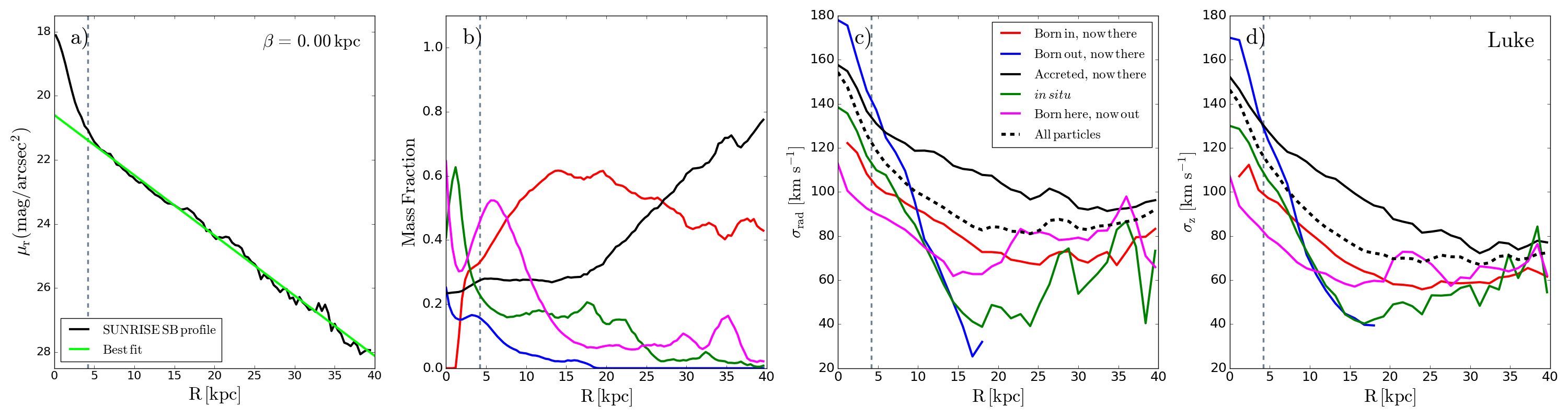} \\
\includegraphics[width=0.95\textwidth]{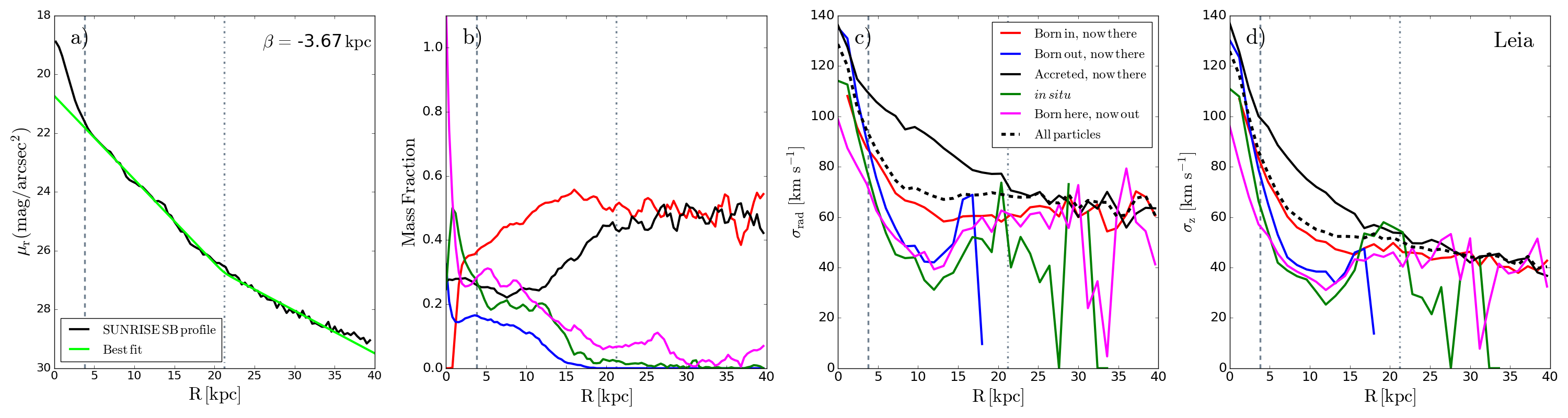} \\
\includegraphics[width=0.95\textwidth]{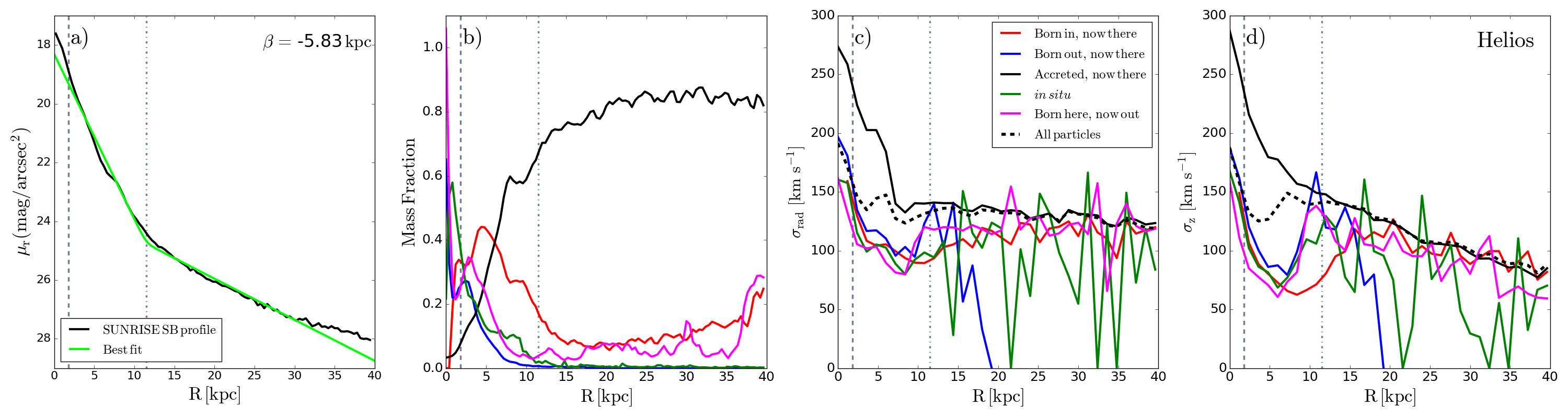} \\
\caption{Same as Fig.~\ref{fractions_1} but for a type I and two type 
III galaxies (with different break strengths), Luke, Leia, and 
Helios.}
\label{fractions_2}
\end{figure*}

Several works seem to suggest that the role of radial redistribution of 
stars could be a decisive agent in shaping the SB profiles of spiral 
galaxies \citep[][]{2008ApJ...675L..65R, 2009ApJ...705L.133M, 
2009MNRAS.398..591S}. In this study we take advantage of the fact that 
the {\tt RaDES} sample presents realistic type I, II, and III galaxies 
to assess if different radial redistribution patterns might give rise to 
the different observed SB profiles. To this aim, we characterise the 
radial redistribution of stars in the whole {\tt RaDES} sample with SB 
profiles ranging from strong type II breaks ($\beta \sim$~5, e.g. 
Castor) to strong type III breaks ($\beta \sim$~-6, e.g. Helios). Such 
characterisation is based on the comparison of radial profiles of the 
mass fraction of stars with some specific characteristics according to 
their birth and current locations (see Figs.~\ref{fractions_1} 
and~\ref{fractions_2}).

On the left-hand panel ({\it a}) of Figs.~\ref{fractions_1} and~\ref{fractions_2} 
we show the SB profiles of six representative examples of galaxies 
displaying different SB profile types and break strengths. The 
second ({\it b}) panel of the same figures shows the radial redistribution of 
stars in those galaxies; to that aim, this panel show the 
radial profiles of the mass fraction of different stellar 
sub-populations according to their birth and current locations. The 
different sub-populations under analysis are: i) stars that are 
currently located at given galactocentric distances (denoted by $\rm R$) 
coming from the inner parts (outward-moving stars, red line); ii) stars 
that are currently located at different galactocentric distances ($\rm 
R$) coming from the outer parts (inward-moving stars, blue line); iii) 
``in situ'' stars (stars that have not move from their birth place, 
green line); iv) stars born there and now located at a larger radius 
(magenta line); and v) accreted stars (black line). Although some of the 
above sub-populations definitions are self-explanatory, ``in situ'' and 
accreted star require precise definition that could potentially affect 
our results. We define accreted stars as those whose R$_\mathrm{birth} > 
$20 kpc and $|\rm z_{\rm birth}| > $~3 kpc 
\citep[following][]{2016A&A...586A.112R}. We consider ``in situ'' stars as those 
whose absolute value of R$_\mathrm{birth} - $R$_\mathrm{current}$ is 
less than 0.2 disc scale-lengths (see Sect.~\ref{SB_prof_char}). In the 
subsequent analysis we will distinguish, not only among galaxies with 
different SB profiles (I, II, or III), but we will also consider the 
fact that {\tt RaDES} type II galaxies have a more external upbending 
profile (type II breaks denoted by the gray dotted-dashed vertical lines 
while type III breaks by the gray dotted vertical lines). 

According to Figs.~\ref{fractions_1} and~\ref{fractions_2} we can claim 
that all galaxies have similar characteristics (regardless of their SB 
profile) in the bulge region\footnote{The bulge region is defined along 
this work as the inner part where the difference between the observed 
light profile and the best fit of the disc light is higher than 0.1 mag 
arcsec$^{-1}$}, delimited by the black vertical dashed line. Roughly 
speaking (and valid for most of the systems), the very centre is 
characterised by stars coming from regions at larger radius (blue line) 
but it is particularly noticeable the amount of stars born in the 
very centre that have moved outward (magenta line). As we move outward, 
the amount of ``in situ'' stars (green line) rises to a peak around the 
middle of the bulge-dominated region (closer to the centre). From the 
location of this peak until the boundary between the bulge and the disc, 
the fraction of stars coming from the inner regions (red line) 
increases. Accreted stars in this central region do not show a common 
behaviour in all the galaxies, but display behaviours ranging from 
domination of accreted stars (e.g. Ben), to those where the amount of 
accreted stars is negligible (Helios or Selene).

It is in the disc-dominated region (beyond the bulge-dominated part of 
the galaxy) where discrepancies are larger between type I, II, and III 
galaxies. The region of the disc prior to the break radius for type II 
galaxies is mainly dominated by stars coming from the inner regions (red 
line) with the peak located at the break radius or beyond. It is worth 
noticing that the region where the disc starts to dominate (inner disc, 
before the break) presents an exodus of stars towards the outer regions 
(magenta line) that is slightly more evident for galaxies with strong 
rather than weaker type II breaks. This exodus of stars plays an 
important role in shaping type II breaks (as we demonstrate in 
Sect.~\ref{facts}). Beyond the type II break, the more important stellar 
sub-populations that we find are accreted stars (black line) and stars 
from the inner regions (red line), with the former gaining in importance 
as we move towards weaker breaks (Castor to Ben) and as we move 
outwards. All type II galaxies from {\tt RaDES} are characterised by an 
excess of light (type III secondary break) in the outer parts. The 
excess of light in these outermost regions is the result of outward 
motions (red line) and satellite accretion (black line), i.e. the accumulation 
of outward and accreted stars. ``In situ'' 
stars (green line) and stars from the outer parts (blue line) are found 
throughout the disc region prior to the type III secondary break (more 
important for strong type II breaks), with a decrement in the mass 
fraction of both sub-populations from the position of the type II break 
outwards.

However, there is a change in the behaviour of the ``in situ'' stars 
(green line) and stars from outer parts (blue line) for type I and III 
galaxies with respect to the type II systems. The fraction of stars in 
these 2 sub-populations is lower across the entire disc (after the 
bulge-dominated region) for type I and III galaxies than for type II 
systems (see Sect.~\ref{facts} where we further quantify this 
statement). Regarding these two types of galaxies (I and III), the main difference 
displayed between them is found in the radial distribution of accreted 
stars (black line) and stars coming from the inner parts (red line). At 
intermediate radii outward-moving stars (red line) dominate, while at 
larger galactocentric distances accreted stars are the main component 
for type I galaxies. The importance of accreted stars is higher for type 
III galaxies with respect to type I systems, especially if they display 
strong type III breaks (Helios). While type I and weak type III galaxies 
display quite extended distributions (across tens of kpcs), Helios (a 
type III galaxy presenting a strong break) presents a concentration of 
outward-moving stars right after the bulge dominated region, probably as 
a consequence of the huge amount of accreted stars in those parts.

The middle-right and right-hand panels ({\it c} and {\it d}) of 
Figs.~\ref{fractions_1} and~\ref{fractions_2} show the radial ({\it c}) 
and vertical ({\it d}) velocity dispersion profiles for the different 
subpopulations analysed. In this case, we have decided to add the profiles 
considering all the stellar particles in the simulation (black dashed lines, current locations). 
It is clear that the inner and outer velocity dispersion profiles display 
distinct behaviours. All galaxies (regardless of their SB profiles) 
exhibit the highest values of the velocity dispersion (radial and vertical) 
in the centre, followed by a radial decrement along the inner regions, i.e. from  
the bulge until a location around the main break. 

In contrast, the outer parts may display either a flattening in the dispersion profiles 
at larger radii or an upturn depending on the galaxy SB profile.
Type II galaxies exhibit outer velocity dispersion profiles for ``in situ'' stars (green) 
that are ``U-shaped'', with the highest values found in the type III part of the profile 
and the minimum generally located beyond the type II break but before the type III one. 
The same behaviour is displayed by outward-moving stars (red) and considering 
all stellar particles together (dashed black line). Similar ``U-shaped'' profiles are also 
displayed by the ``in situ'' subpopulation in the case of type I galaxies. However, an 
outer flattening (with a slight increase) is displayed by the outward-moving stars and 
this shape is also found if we compute the velocity dispersion profiles considering all 
stellar particles together. On the other hand, outer flattenings or oscillating outer 
profiles are found for all subpopulations in the case of type III galaxies starting slightly 
before the location of the type III break. We must note that 
the reported inner decrement in the velocity dispersion profiles of all subpopulations is steeper for stronger 
type III breaks. On the other hand, the velocity dispersion profiles displayed by accreted 
stars is very similar for all galaxies (including types I, II, and III), namely a gradual radial decrement from the highest 
values (at the centre) to the lowest ones (at around the break location) followed by a flattening 
or a smoother decrement (in the outskirts). The extension of the region dominated by the 
gradual decrement for the velocity dispersion of the accreted stars decreases as we move 
from type II galaxies to type I and III.

The fact that the boundary between both regimes (inner and outer behaviour) seems to be located near the break location 
may suggest a common origin between the light and the velocity dispersion profiles as well as 
several processes working at once.

\subsection{Interpretation}
\label{scenario}

These findings can be interpreted in such way that radial redistribution 
of material (regardless of what physically causes it and including 
accretion) can explain the different SB profiles. In such scenario, the 
presence and strength ($\beta$) of type II breaks are a direct 
consequence of the combination of the radial redistribution of stars and 
accretion. Stars moving outwards produce a deficit of stars in the disc 
region (right after the bulge-dominated zone) before the break radius 
(see magenta line) and, as a consequence, large inner disc scale-lengths 
are found in contrast to the outer part (type II break). As this deficit 
becomes less important and more stars are found in the region 
inmediately located after the bulge region, shorter disc scale-lengths 
are displayed, i.e. weaker breaks (see trend from Castor to Ben).  
However, the formation of a type II break does not only depends on the 
distribution of outward-moving stars; it is the complex combination of 
this distribution and the distribution of accreted stars what finally 
shapes type II galaxies and determine the position of the break. The 
region where type II breaks are found and beyond is again dominated by  
outward-moving stars (solid red line) and accreted stars (black line) 
that are accumulated in that region. 
As the fraction of accreted stars is more important at these 
intermediate galactocentric distances (and also before the type II 
break) we detect weaker type II breaks. Accreted stars smoothen the 
light profiles giving rise to weak type II breaks. The secondary type 
III breaks that are found in type II galaxies are basically formed 
because of the accumulation of accreted and outward-moving (especially 
accreted) stars in these outer parts which produces an outer excess of 
light.

Despite the tendency of outward-moving and accreted stars to dominate the mass 
fraction for all galaxies, it is for type I and III galaxies where this 
superiority is even more important. The fact that the fraction of 
inward-moving and ``in situ'' stars is lower in the case of type I and 
III galaxies make radial redistribution a more important agent in 
shaping SB profiles in these galaxies than in type II galaxies. 
Therefore, it is the mixing induced by outward-moving and, especially, 
accreted stars what causes type II breaks to form pure exponential 
profiles. The final profile (I or III) will depend on the relative 
fraction of outward-moving and accreted stars. As accreted stars start 
to dominate, especially in the outer parts, type III breaks appear and 
become stronger. The overpopulation of outward-moving (red line) stars 
right after the bulge-dominated region found in type III galaxies with 
strong breaks (Helios) make the inner disc scale-length shorter (i.e. a 
steeper inner light profile).

We must warn the reader here that this interpretation does not necessary 
establish a time evolution from type II galaxies to type III systems. We 
are only trying to obtain patterns in the radial redistribution of 
material that can link the different observed SB profiles. Those 
patterns suggest that radial redistribution of material can change the 
SB profile of a galaxy from type II, to I, and finally to type III in 
favourable conditions. However, this interpretation is qualitative and 
needs some quantification to be more robust. In Sect.~\ref{facts} we 
quantify some of the previous analysis.

\subsection{Facts favouring this interpretation}
\label{facts}

If the previous qualitative reasoning is correct and radial motions do 
play a role in shaping SB profiles, then Luke (a galaxy displaying a 
pure exponential light profile) should display a type III break in the 
region located between 25 and 35 kpc attending to Fig.~\ref{fractions_2} 
(upper panel), where accreted stars begin to dominate. The lack of such 
a break in Luke suggest that, although the radial redistribution of 
material accumulating in the outer parts in this galaxy is also 
important, there must be some peculiarities swamping out any possible 
type III breaks. In the following we consider all the systems in {\tt 
RaDES} and quantify radial redistribution as a function of the break 
strength ($\beta$), a continuous quantity, to distinguish between SB 
types, not focusing on individual galaxies as previously done. 
This will allow us to further investigate the reasons for the 
appearance of type I galaxies (and other SB profiles) as well as further 
support our previous interpretation. This analysis of the radial 
redistribution as a function of $\beta$ allows us to assess the effect 
of radial redistribution, not only in shaping type I, II, or III SB 
profiles, but also in shaping the strength of the different breaks. For 
simplicity, we will focus in two regions that will be named {\it inner} 
region and {\it outer} region. The {\it inner } region corresponds to 
the disc-dominated region before the break (type II or III) or before 3 
disc scale-lengths for type I galaxies. The {\it outer} region will be 
the region between the type II break and the type III break for type II 
galaxies, the region beyond the break for type III galaxies, and the 
region beyond 3 disc scale-lengths for type I galaxies. We have decided 
to use 3 disc scale-lengths as the separation radius as it is there 
where the breaks are usually found in real galaxies 
\citep[][]{2006A&A...454..759P, 2008AJ....135...20E}. These two regions 
are key locations in which we can focus the analysis in order to check 
the validity of the previous interpretation.

\begin{figure*}
\centering
\includegraphics[width=0.8\textwidth]{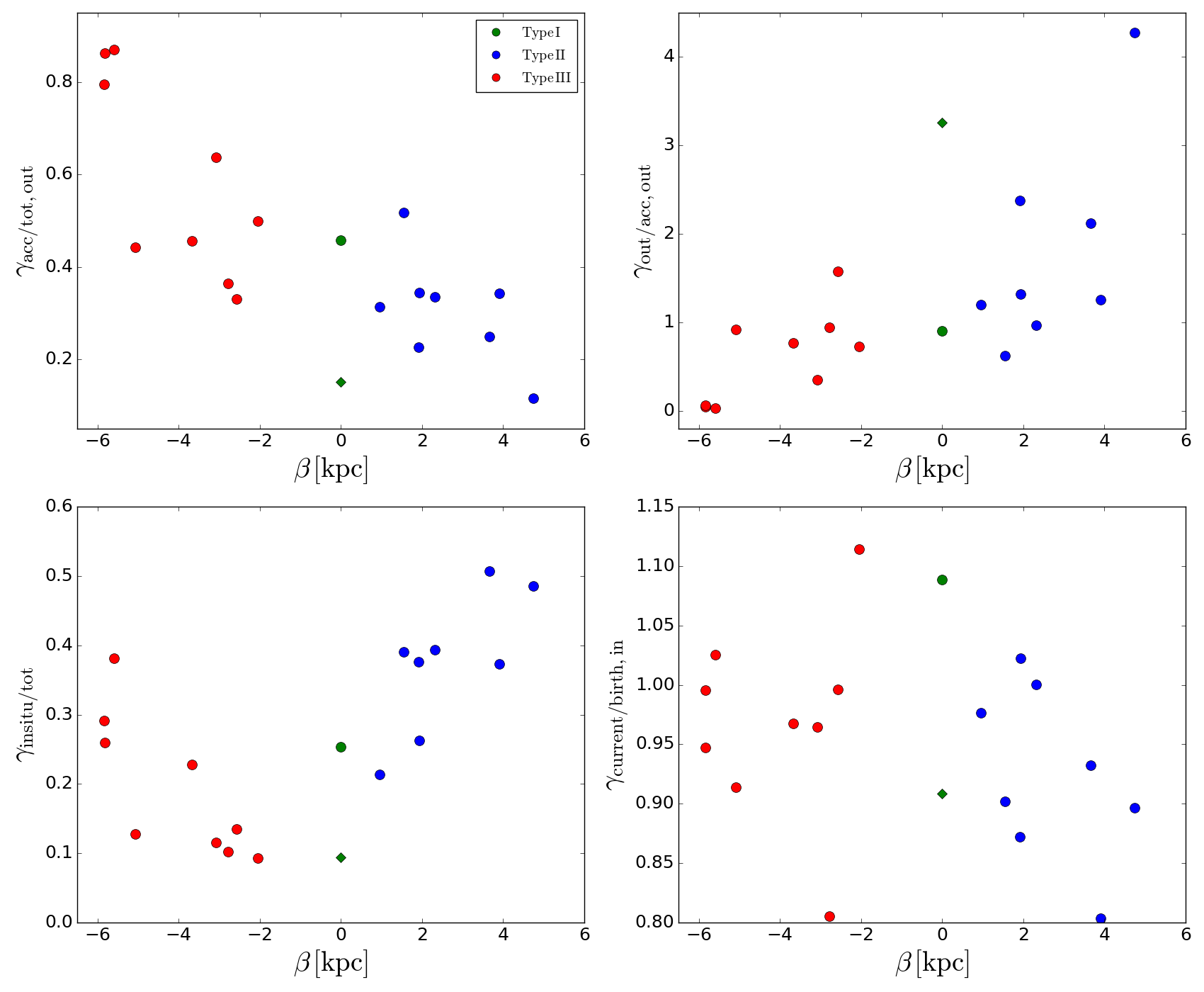}
\caption{Quantification of the radial redistribution as a function of 
the break strength ($\beta$), defined as h$_{\rm in}$ - h$_{\rm out}$. 
Left-hand, top panel: mass fraction of accreted stars currently 
located in the {\it outer} region with respect to the total stellar 
mass in that region ($\rm \gamma_{acc/tot, out}$) as a function of 
$\beta$. Right-hand, top panel: mass fraction of stars coming from 
inner radius with respect to the stellar mass of accreted stars 
currently located in the {\it outer} region ($\rm \gamma_{out/acc, 
out}$) as a function of $\beta$. Left-hand, bottom panel: mass 
fraction of ``in situ'' stars in the whole galaxy with respect to the 
total mass of the galaxy ($\rm \gamma_{insitu/tot}$) as a function of 
$\beta$. Right-hand, bottom panel: mass fraction of stars currently 
located in the {\it inner} region with respect to the stars born in 
the same {\it inner} region ($\rm \gamma_{current/birth, in}$). Green 
points represent type I galaxies, blue points are type II galaxies 
and type III systems are denoted by red points. Zeus is represented by 
a green diamond as it is a type I galaxy with signs of ongoing 
interaction. See text for details.}
\label{beta}
\end{figure*}

In the left-hand, top panel of Fig.~\ref{beta} we represent the mass 
fraction of accreted stars currently located in the {\it outer} region 
with respect to the total stellar mass in the same region ($\rm 
\gamma_{acc/tot, out}$) as a function of $\beta$. We find a clear 
correlation with strong type II galaxies having the lowest values of 
$\rm \gamma_{acc/tot, out}$ and strong type III galaxies displaying the 
highest ones. In the right-hand, top panel of Fig.~\ref{beta} we 
represent the mass fraction of stars coming from inner radius 
(outward-moving stars) with respect to the stellar mass of accreted 
stars in the {\it outer} region ($\rm \gamma_{out/acc, out}$) for the 
{\tt RaDES} galaxies. In this case, type II galaxies with strong breaks 
display the highest values of $\rm \gamma_{out/acc, out}$ while type III 
galaxies with strong breaks displays the lowest ones. The mass fraction 
of ``in situ'' stars (along the entire galaxy) with respect to the total 
galaxy mass ($\rm \gamma_{insitu/tot}$) is represented in the left-hand, 
bottom panel. In this case we find a ``U-shaped'' relation between both 
magnitudes with type II galaxies displaying the highest values and weak 
type III systems showing the lowest. Galaxies with strong type III 
breaks also display a good amount of ``in situ'' stars (comparable with 
systems with weak type II breaks). Finally, in the right-hand, bottom 
panel we show the variation of the mass fraction of stars currently 
located in the {\it inner} region with respect to the mass of stars 
stars born in the same {\it inner} region ($\rm \gamma_{current/birth, 
in}$) as a function of $\beta$. Type II galaxies exhibit an interesting 
correlation with some scatter in which strong type II breaks display low 
values of $\rm \gamma_{current/birth, in}$ and weak type II breaks or 
even type I systems present higher values.

The four correlations support the proposed scenario in which different 
degrees of radial redistribution of material give rise to the different 
observed SB profiles. The correlation found for $\rm \gamma_{acc/tot, 
out}$ reflects that accretion is more important in type I and III 
galaxies than in type II galaxies, confirming that accretion can convert 
type II breaks into pure exponentials or even type III galaxies. The 
behaviour of $\rm \gamma_{out/acc, out}$ with type II galaxies 
displaying larger values reflects again that accreted stars are less 
abundant in the {\it outer} region for these galaxies (causing outer 
disc scale-length to be shorter) and that outward-moving stars 
accumulate around the type II break and beyond. The shape of the $\rm 
\gamma_{insitu/tot}$ vs. $\beta$ relation allows us to confirm that the 
amount of ``in situ'' stars for type II galaxies is higher than for type 
I or III galaxies, verifying our interpretation that radial 
redistribution being less important in type II galaxies. However, the 
three type III galaxies with the strongest breaks invert this trend and 
display similar values of weak type II discs. It is the vast amount of 
accreted stars in these systems (see the upper left-hand panel of 
Fig.~\ref{beta}) that causes them to have such strong type III breaks 
despite the quantity of ``in situ'' stars. It is the correlation of $\rm 
\gamma_{current/birth, in}$ with $\beta$ for type II galaxies that leads 
us to the conclusion that one of the main factors shaping type II breaks 
is the exodus of stars from {\it inner} regions. The lack of stars in 
the {\it inner} region produces a shallower inner light profiles which 
strengthens the break. The fact that this behaviour is not found for 
type III galaxies reinforces the interpretation provided in this work.

It is worth noticing that the importance of all the processes described 
is always intermediate for type I galaxies with respect to the other two 
types. This allows us to conclude that type I galaxies are ``boundary'' 
galaxies with properties in between those of type II and type III 
galaxies and that they exhibit small peculiarities (that need to be 
further analysed) swamping out any possible breaks. As the reader may 
have noticed, in the four correlations there is always one type I 
galaxy slightly off the correlation, Zeus (represented in 
Fig.~\ref{beta} with green diamonds). Zeus still has a satellite 
companion and thus it is not the best example for this analysis. The 
presence of the companion in Zeus explains its position out of the 
correlation in all the relations \citep[see][for further information on 
Zeus]{2012A&A...547A..63F}.

Coming back to Luke, the lack of a type III break can be explained based 
on the radial redistribution of material in these outer regions. 
Although accreted stars dominate with respect to the other 
subpopulations (black line), it is not as important in absolute terms as 
for the rest of type III galaxies (see top-right panel of 
Fig.~\ref{beta}). In addition, there is a clear negative radial trend of 
the stars coming from the inner regions that end up at such 
galactocentric regions (red line). Both facts, along with the exodus of 
stars located at those external parts and that have moved outwards 
(magenta line), make that stars do not accumulate in the outskirts 
preventing any type III break to be observed.

\subsection{On the behaviour of the velocity dispersion radial profiles}
\label{vel_disp}

Previous works have found that type III profiles come from a kinematically hot 
component. \citet[][]{2017MNRAS.470.4941H} suggested that a combination of a strong 
bar with a low initial halo spin is able to make stars migrate outward to form 
type III breaks that are populated by stars that acquire high velocity dispersions 
in the migration process. In addition, another way to inhabit the outer parts of type II 
galaxies with a hot component was found by \citet[][]{2012A&A...548A.126M} via smooth accretion 
of gas. In that work, this accretion leads to the formation of outer type III breaks coexisting with the former downbending 
profile. The resemblance between their type-II+type-III combined profiles with the ones 
shown in this work is remarkable. However, while \citet[][]{2012A&A...548A.126M} found a 
clear correlation between the location of this secondary type III break and an upturn in 
the stellar radial velocity dispersion (see their figure 17), we do not find such clear 
relation. We obtain that the stellar velocity dispersion profiles of the type II {\tt RaDES} galaxies display 
upturns starting at intermediate locations between the inner type II break and the outer type III one (e.g. Castor) 
or related with the location of the primary type II break (e.g. Selene and Ben). 
The discrepancies among both works are understandable considering the differences in the 
simulation recipes as well as the fact that the \citet[][]{2012A&A...548A.126M} analysis do 
not include stellar accretion. In fact, the location of the upturn in velocity dispersion seems to be related 
(in all cases) with the position where the mass fraction of accreted stars (see panel b) starts to 
increase, suggesting that accreted stars have a greater impact on the global velocity dispersion than on the 
light profile. However, not only the ``global'' radial and vertical velocity 
dispersion profiles (considering all particles, including accretion) show this upturn, but also 
other subpopulations of non-accreted stars (see Figs.~\ref{fractions_1} and~\ref{fractions_2}).
Regarding pure type III galaxies, they do not display this velocity dispersion upturns 
at all. In contrast, they mainly show flattenings that start generally before the type III 
breaks, again related with the radial position where the accreted stars begin to dominate. 

To sum up, we can claim that all the {\tt RaDES} galaxies display stellar velocity dispersions 
in the outer parts that are higher than that expected from the extrapolation of the behaviour 
in the inner parts (before the regions where the breaks appear), although only type I and II galaxies 
exhibit clear upturns that might be related with the accretion.

\section{Effect of radial redistribution in shaping stellar population 
age and metallicity profiles}
\label{result2}

According to the previous section, radial redistribution of stars seems 
to play an important role in shaping the different observed SB profiles. 
However, from an observational point of view, this statement is 
difficult to confirm and might fall into the box of the speculative 
theories. To solve this issue, other observables have to be analysed in 
order to provide this hypothesis with predictions that can be tested. 
Some of the observables that can be computed using modern observing 
techniques are stellar age and metallicity profiles, especially in the 
inner regions where the signal is higher. 
\citet[][]{2017A&A...604A...4R} (hereafter TRL17) recently studied such 
profiles using the spectroscopic data provided by the CALIFA survey 
\citep[][]{2012A&A...538A...8S}. The authors, analysing a sample of 214 
spiral galaxies, found that type II systems 
tend to display steeper stellar age and metallicity profiles than type I 
and III galaxies, with the latter showing the shallowest ones. They 
interpret those findings as a consequence of radial redistribution of 
material being more effective in type III galaxies than in type I or II 
systems, in agreement with the scenario proposed in this work.

\begin{figure*}
\centering
\includegraphics[width=0.8\textwidth]{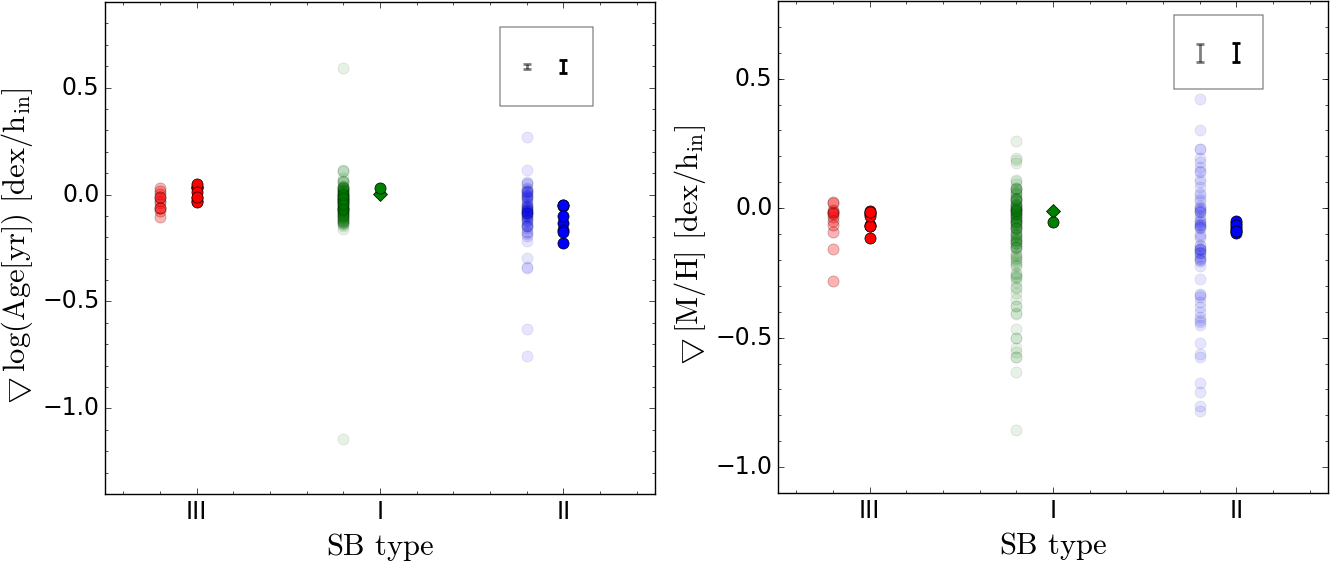} \\
\caption{Comparison of the distributions of the inner gradients of the 
stellar age (left-hand panel) and metallicity (right-hand panel) 
profiles for type I (green), II (blue), and III (red) galaxies from 
{\tt RaDES} with the CALIFA mass-weighted distributions (TRL17, 
symbols with transparency). Zeus is again represented by a green diamond (type I galaxy 
with signs of ongoing interaction). Typical errors in the gradient 
values for the observations and the simulations are shown with the 
black error bars within the boxes (left, typical errors in the CALIFA values; 
right, typical errors in the case of the gradients from the simulations).}
\label{rades_ste_pop_grads}
\end{figure*}

In order to compare with those observational results, we compute 
mass-weighted age and metallicity profiles for all the {\tt RaDES} 
galaxies by averaging all disc particles using 0.5\,kpc-wide radial bins 
\citep[see][]{2016A&A...586A.112R}. Afterwards, we perform error 
weighted linear fits to those profiles in the inner and the outer 
regions, separately. We define these inner and outer regions following 
TRL17 prescription and matching the inner and outer regions defined 
previously (see Sect.~\ref{facts}). We decided to perform this analysis 
only in the inner parts of the profiles as in the analysis presented in 
TRL17. In Fig.~\ref{rades_ste_pop_grads} we show the distributions of 
the inner gradients for the stellar age (left-hand panel) and 
metallicity (right-hand panel) profiles distinguishing between type I 
(green), II (blue), and III (red) galaxies over-plotting the results 
coming from the CALIFA observations (TRL17, symbols with transparency). 
Table~\ref{tab:av_gradients} shows the error-weighted average and 
dispersion values of the inner gradients of the stellar age and 
metallicity for the type I, II, and III galaxies from {\tt RaDES}.

The values of the age and metallicity gradients for individual galaxies 
from the simulations are 
consistent with the values presented in TRL17 for the CALIFA sample (see 
Fig.~\ref{rades_ste_pop_grads}). The low number of simulated galaxies 
with respect to the observed ones hampers a complete comparison of both 
distributions, however, some similarities can be highlighted. Type II 
galaxies display a larger dispersion in age gradient for the simulations 
and the observations while in the case of metallicity gradients the 
dispersions are compatible for the three types of galaxies. Type II 
galaxies present the steepest average gradients for the age (-0.09 
dex/h$_{\rm in}$) and the metallicity (-0.07 dex/h$_{\rm in}^{-1}$). On 
the other hand, type I systems show the shallowest profiles (0.004 and 
-0.02 dex/h$_{\rm in}$ for age and metallicity, respectively) with type 
III systems displaying intermediate values (or even positive values in 
the case of the age: 0.02 and -0.05 dex/h$_{\rm in}$, for the age and 
metallicity gradients, respectively). In TRL17, type III systems are 
found to have the shallowest profiles with a clear trend from type II 
(steepest) to type I (intermediate) and type III. The differences 
between the simulated values presented in this work and the 
observational results from CALIFA are reasonable and easily explained 
considering the low number of simulated systems analysed. However, the 
similarities and the consistency between the gradients for individual galaxies 
are reassuring.

\begin{table*}
\centering
\small
\begin{tabular}{lccc}
\hline
 & \multicolumn{3}{c}{SB profile}\\ \hline
 & Type III & Type I & Type II \\
{\large $ \rm \triangledown$} log(Age[yr]) [dex/h$_{\rm in}$]  &  0.02 (0.03)  & 0.004 (0.01) & -0.09 (0.06) \\
{\large $ \rm \triangledown$} [M/H] [dex/h$_{\rm in}$]    & -0.05 (0.03)  & -0.02 (0.02) & -0.07 (0.02) \\
\hline
\end{tabular}
\caption[Stellar population inner gradients (averaged 
values)]{Error-weighted average and dispersion values of the inner 
gradients of the stellar age and metallicity (see text for details) 
for the {\tt RaDES} set of galaxies. Units are in dex/h$_{\rm in}$.}
\label{tab:av_gradients}
\end{table*}

\begin{figure*}
\centering
\includegraphics[width=0.8\textwidth]{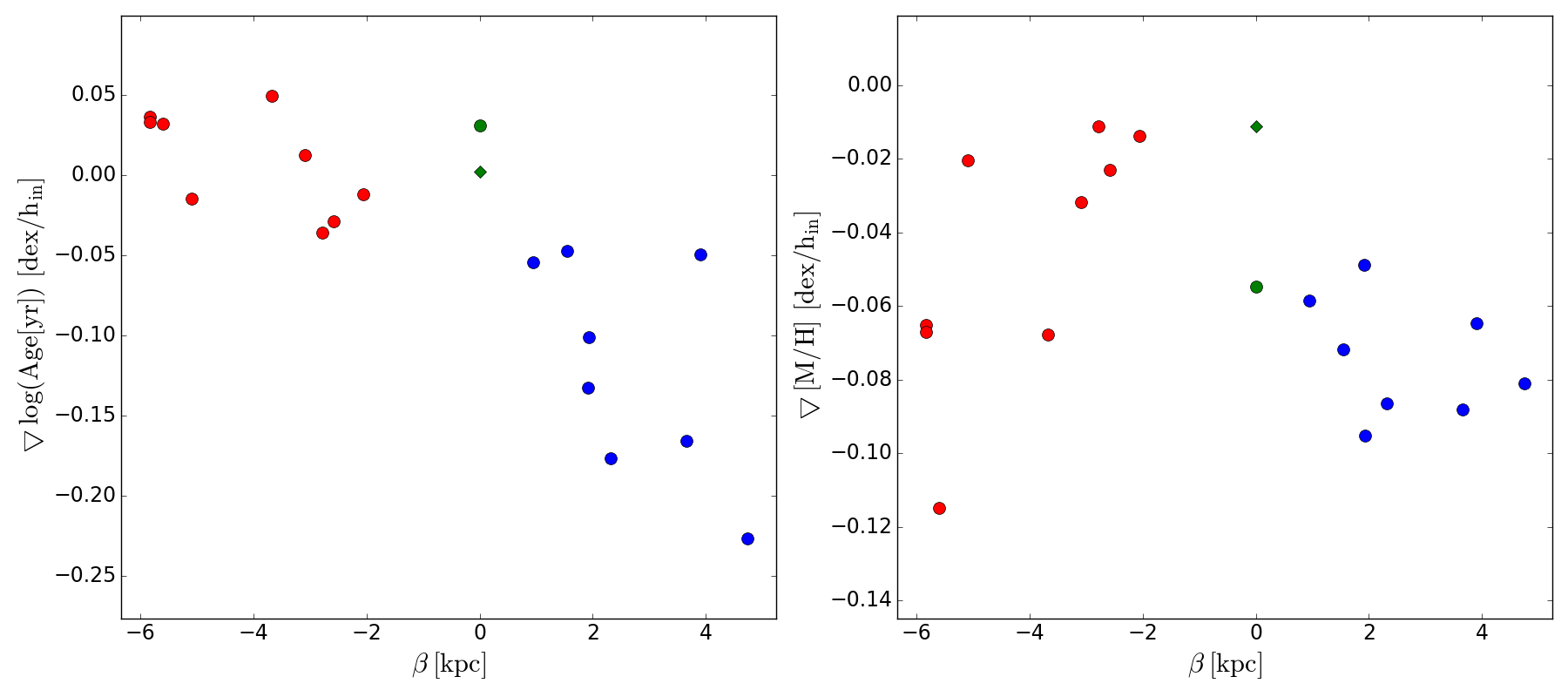} \\
\caption{Stellar age (left-hand panel) and metallicity (right-hand 
panel) inner gradients as a function of the break strength ($\beta$) 
for the {\tt RaDES} set of galaxies. The gradients for type I, II, and 
III galaxies are indicated in green, blue, and red, respectively. Zeus 
is again represented by a green diamond (type I galaxy with signs of 
ongoing interaction).}
\label{rades_ste_pop_grads_beta}
\end{figure*}

In Fig.~\ref{rades_ste_pop_grads_beta} we study the dependency of the 
stellar age and metallicity gradients displayed by the {\tt RaDES} 
galaxies as a function of the strength of their breaks ($\beta$). If 
radial redistribution of material does play an important role in shaping 
SB profiles (see Sect.~\ref{scenario}) and it also affects age or 
metallicity gradients of the inner disc (see 
Fig.~\ref{rades_ste_pop_grads}), then there should be some kind of 
correlation between these gradients and the detailed shape of the break. 
In fact, this is what we show in Fig.~\ref{rades_ste_pop_grads_beta}. 
Type II galaxies with strong breaks (less influenced by radial 
redistribution, especially accretion) display steeper stellar age and 
metallicity profiles than type II systems with weak breaks or even type 
I galaxies. A consistent behaviour is also found in the case of the age 
profiles for type III galaxies in which systems with weak breaks (less 
affected by accretion) display steeper profiles than galaxies with 
strong breaks, in which case, even positive gradients are displayed. 
However, in the case of the metallicity gradients this trend is 
inverted. Type III galaxies with weak breaks have shallower profiles 
than systems with strong breaks (which are more affected by accretion). 
This inverted trend can be explained by the characteristics of the 
accreted stars. In \citet[][]{2016A&A...586A.112R} we showed that 
accreted stars populating the {\tt RaDES} galaxies are mainly old, 
metal-poor stars. As a consequence, the accumulation of accreted stars 
in the outer parts of type III galaxies giving rise to these kind of 
systems causes the outer parts to be old (positive age gradients) and 
relatively metal-poor (producing steeper negative profiles). Based on 
Fig.~\ref{rades_ste_pop_grads_beta} we can conclude that, not only does 
radial redistribution of material affect the shape of observed light 
distributions, but also that accretion has to play a key role in shaping 
observed stellar age and metallicity profiles.
 
\begin{figure*}
\centering
\includegraphics[width=0.95\textwidth]{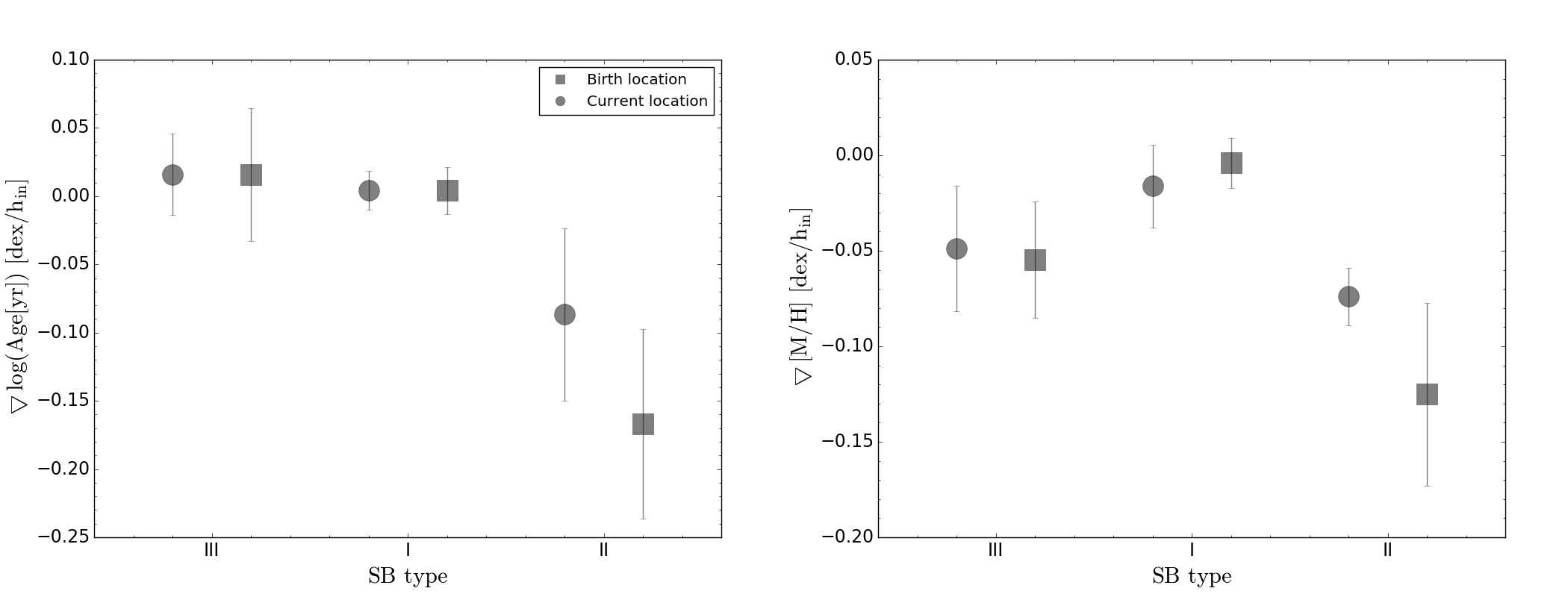} \\
\caption{Distribution of the inner gradients of the stellar age 
(left-hand panel) and metallicity (right-hand panel) profiles for type 
I, II, and III galaxies considering current locations (grey circles) 
or birth locations (grey squares) of the stars to compute the stellar 
age and metallicity profiles.}
\label{new_test_patri}
\end{figure*}

All these results and the similarities between the observational results 
(TRL17) and the theoretical work presented in this paper seem to 
indicate that radial redistribution of material and accretion shape the 
SB profiles and generally flatten the age and the metallicity profiles 
with the exception of the metallicity profiles for type III galaxies due 
to the accumulation of metal-poor, accreted stars in the outer disc. The 
greater the efficiency of this redistribution, the larger the flattening 
effect would be. As a consequence, we would expect steeper age and 
metallicity gradients at birth that what we observe presently due to the 
effect of the radial redistribution of material. To shed light onto this 
aspect and the true role of this redistribution in shaping the age and 
metallicity profiles we have also analysed the inner gradients of these 
stellar parameters in the absence of radial motions, i.e. considering 
birth locations (see Fig.~\ref{new_test_patri}).

Figure~\ref{new_test_patri} shows the same information as 
Fig.~\ref{rades_ste_pop_grads} but focusing on the effect of radial 
motion of stars on the stellar age and metallicity gradients. In this 
case we show mean values for type I, II and III systems, not the values 
for all the individual galaxies. Grey circles represent the mean values 
of the stellar age and metallicity gradients (averaging among SB types) 
considering stars located at their current locations. Grey squares 
represent the same mean quantities but this time computed using stars 
located at their birth locations (thus avoiding the effect of radial 
motions). Strikingly, the inner stellar parameter gradients for type I 
and III galaxies do not seem to be specially affected by radial 
redistribution of material despite steeper gradients being expected when 
considering birth locations and shallower ones at the current locations. 
The case of type II galaxies however shows that the migration of stars 
flattens the stellar profiles considerably. The fact that, even at birth 
locations type I and III galaxies seem to display shallow age and 
metallicity profiles (and definitely shallower than type II galaxies), 
suggests that the differences found in Fig.~\ref{rades_ste_pop_grads} must 
have been imprinted at birth. This points toward the existence of formation 
mechanisms producing steep stellar age and metallicity profiles for type 
II galaxies and shallow profiles for type I and III systems. These 
findings deserve further investigation focusing on the early stages of 
the formation of these systems that is beyond the scope of this paper, 
focused on the relation between radial redistribution and the 
shape of the SB profiles.

\section{Conclusions}
\label{conclusions}

In this work we propose a scenario in which radial redistribution of 
stars can significantly affect the light distributions shaping observed 
SB profiles. Redistribution of material, especially outward-moving and 
accreted stars, is important in all {\tt RaDES} galaxies regardless of 
their SB profile. The accretion and migration of stars particularly 
prevails in type I and III galaxies while the fraction of ``in situ'' 
stars is slightly higher in type II systems. Both aspects suggest that 
radial redistribution is less significant in systems with downbending 
light profiles. Outward-moving stars from the region just beyond the 
bulge produce extended inner light profiles and the accumulation of 
these and accreted stars determine the presence and strength of type II 
breaks. An increase in the fraction of accreted stars causes the fading 
of the type II breaks into pure exponentials and even the shaping of 
type III breaks whose strength depends on the amount of accreted stars 
settled in the outer parts. We have proven that type I galaxies present 
properties that lie between those of type II and type III galaxies.
In addition, the stars populating the locations where 
breaks are found and beyond (where accreted stars begin to dominate) present higher velocity dispersions than that 
expected by extrapolating the behaviour in the inner parts in agreement with 
previous works \citep[e.g.][]{2012A&A...548A.126M, 2017MNRAS.470.4941H}. This 
scenario seems to leave signatures in the radial distribution of stellar 
properties such as age and metallicity, especially due to the effect of 
accreted stars. Shallower inner profiles are found in galaxies where 
radial redistribution is more important (type I and III) while type II 
galaxies present steeper inner gradients. These signatures have been 
observed recently in a sample of nearby galaxies from the CALIFA survey 
\citep[][]{2017A&A...604A...4R}. However, these differences in the inner 
gradients of stellar population properties can be also found in the 
absence of stellar radial motion suggesting that they may be imprinted 
at birth. Further work merging observational analysis with realistic cosmological simulations\footnote{Future 
analyses will aim to identify star particles more rigosourly accordingly to their 
observational characteristics, as proposed in \citet[][]{2017arXiv170901523T}.} 
in a larger number of systems to improve the statistics is needed to properly 
understand how the different SB profiles come into shape.

\begin{acknowledgements}
We thank the referee for very useful suggestions and comments that have helped improve the current version of this manuscript. 
This research has been partly supported by the Spanish Ministry of 
Science and Innovation (MICINN) under grants AYA2014-53506-P and AYA2014-56795-P; 
and by the Junta de Andaluc\'ia (FQM-108). We 
acknowledge the generous allocation of resources from the Partnership 
for Advanced Computing in Europe (PRACE) via the DEISA Extreme 
Computing Initiative (PRACE-3IP Project RI-312763 and PRACE-4IP 
Project 653838), STFC's DiRAC Facility (COSMOS: Galactic Archaeology - 
ST/J005673/1, ST/H008586/1, ST/K00333X/1), and the University of 
Hull's High Performance Computing Facility (\textsc{viper}). TRL 
thanks the support of the Spanish Ministerio de Educaci\'on, Cultura y 
Deporte by means of the FPU fellowship. CGF acknowledges support from 
the FP7 European Research Council Starting Grant LOCALSTAR.
\end{acknowledgements}

\bibliographystyle{aa}       
\bibliography{bibliography}  

\begin{thebibliography}{48}
\expandafter\ifx\csname natexlab\endcsname\relax\def\natexlab#1{#1}\fi

\bibitem[{{Abadi} {et~al.}(2003){Abadi}, {Navarro}, {Steinmetz}, \&
  {Eke}}]{2003ApJ...591..499A}
{Abadi}, M.~G., {Navarro}, J.~F., {Steinmetz}, M., \& {Eke}, V.~R. 2003, \apj,
  591, 499

\bibitem[{{Bakos} {et~al.}(2008){Bakos}, {Trujillo}, \&
  {Pohlen}}]{2008ApJ...683L.103B}
{Bakos}, J., {Trujillo}, I., \& {Pohlen}, M. 2008, \apjl, 683, L103

\bibitem[{{Bernard} {et~al.}(2012){Bernard}, {Ferguson}, {Barker}, {Hidalgo},
  {Ibata}, {Irwin}, {Lewis}, {McConnachie}, {Monelli}, \&
  {Chapman}}]{2012MNRAS.420.2625B}
{Bernard}, E.~J., {Ferguson}, A.~M.~N., {Barker}, M.~K., {et~al.} 2012, \mnras,
  420, 2625

\bibitem[{{Bernard} {et~al.}(2015){Bernard}, {Ferguson}, {Richardson}, {Irwin},
  {Barker}, {Hidalgo}, {Aparicio}, {Chapman}, {Ibata}, {Lewis}, {McConnachie},
  \& {Tanvir}}]{2015MNRAS.446.2789B}
{Bernard}, E.~J., {Ferguson}, A.~M.~N., {Richardson}, J.~C., {et~al.} 2015,
  \mnras, 446, 2789

\bibitem[{{Bird} {et~al.}(2012){Bird}, {Kazantzidis}, \&
  {Weinberg}}]{2012MNRAS.420..913B}
{Bird}, J.~C., {Kazantzidis}, S., \& {Weinberg}, D.~H. 2012, \mnras, 420, 913

\bibitem[{{Bland-Hawthorn} {et~al.}(2005){Bland-Hawthorn}, {Vlaji{\'c}},
  {Freeman}, \& {Draine}}]{2005ApJ...629..239B}
{Bland-Hawthorn}, J., {Vlaji{\'c}}, M., {Freeman}, K.~C., \& {Draine}, B.~T.
  2005, \apj, 629, 239

\bibitem[{{Borlaff} {et~al.}(2016){Borlaff}, {Eliche-Moral}, {Beckman}, \&
  {Font}}]{2016A&A...591L...7B}
{Borlaff}, A., {Eliche-Moral}, M.~C., {Beckman}, J., \& {Font}, J. 2016, \aap,
  591, L7

\bibitem[{{Elmegreen} \& {Struck}(2016)}]{2016ApJ...830..115E}
{Elmegreen}, B.~G. \& {Struck}, C. 2016, \apj, 830, 115

\bibitem[{{Erwin} {et~al.}(2005){Erwin}, {Beckman}, \&
  {Pohlen}}]{2005ApJ...626L..81E}
{Erwin}, P., {Beckman}, J.~E., \& {Pohlen}, M. 2005, \apjl, 626, L81

\bibitem[{{Erwin} {et~al.}(2008){Erwin}, {Pohlen}, \&
  {Beckman}}]{2008AJ....135...20E}
{Erwin}, P., {Pohlen}, M., \& {Beckman}, J.~E. 2008, \aj, 135, 20

\bibitem[{{Few} {et~al.}(2012){Few}, {Gibson}, {Courty}, {Michel-Dansac},
  {Brook}, \& {Stinson}}]{2012A&A...547A..63F}
{Few}, C.~G., {Gibson}, B.~K., {Courty}, S., {et~al.} 2012, \aap, 547, A63

\bibitem[{{Florido} {et~al.}(2006){Florido}, {Battaner}, {Guijarro},
  {Garz{\'o}n}, \& {Castillo-Morales}}]{2006A&A...455..467F}
{Florido}, E., {Battaner}, E., {Guijarro}, A., {Garz{\'o}n}, F., \&
  {Castillo-Morales}, A. 2006, \aap, 455, 467

\bibitem[{{Freeman}(1970)}]{1970ApJ...160..811F}
{Freeman}, K.~C. 1970, \apj, 160, 811

\bibitem[{{Gibson} {et~al.}(2013){Gibson}, {Courty}, {Cunnama}, \&
  {Moll{\'a}}}]{BuenosAires}
{Gibson}, B.~K., {Courty}, S., {Cunnama}, D., \& {Moll{\'a}}, M. 2013,
  Asociacion Argentina de Astronomia La Plata Argentina Book Series, 4, 57

\bibitem[{{Gogarten} {et~al.}(2010){Gogarten}, {Dalcanton}, {Williams}, {Ro{\v
  s}kar}, {Holtzman}, {Seth}, {Dolphin}, {Weisz}, {Cole}, {Debattista},
  {Gilbert}, {Olsen}, {Skillman}, {de Jong}, {Karachentsev}, \&
  {Quinn}}]{2010ApJ...712..858G}
{Gogarten}, S.~M., {Dalcanton}, J.~J., {Williams}, B.~F., {et~al.} 2010, \apj,
  712, 858

\bibitem[{{Grand} {et~al.}(2016){Grand}, {Springel}, {Kawata}, {Minchev},
  {S{\'a}nchez-Bl{\'a}zquez}, {G{\'o}mez}, {Marinacci}, {Pakmor}, \&
  {Campbell}}]{2016MNRAS.460L..94G}
{Grand}, R.~J.~J., {Springel}, V., {Kawata}, D., {et~al.} 2016, \mnras, 460,
  L94

\bibitem[{{Herpich} {et~al.}(2015){Herpich}, {Stinson}, {Dutton}, {Rix},
  {Martig}, {Ro{\v s}kar}, {Macci{\`o}}, {Quinn}, \&
  {Wadsley}}]{2015MNRAS.448L..99H}
{Herpich}, J., {Stinson}, G.~S., {Dutton}, A.~A., {et~al.} 2015, \mnras, 448,
  L99

\bibitem[{{Herpich} {et~al.}(2017){Herpich}, {Stinson}, {Rix}, {Martig}, \&
  {Dutton}}]{2017MNRAS.470.4941H}
{Herpich}, J., {Stinson}, G.~S., {Rix}, H.-W., {Martig}, M., \& {Dutton}, A.~A.
  2017, \mnras, 470, 4941

\bibitem[{{Jonsson}(2006)}]{2006MNRAS.372....2J}
{Jonsson}, P. 2006, \mnras, 372, 2

\bibitem[{{Leitherer} {et~al.}(1999){Leitherer}, {Schaerer}, {Goldader},
  {Gonz{\'a}lez Delgado}, {Robert}, {Kune}, {de Mello}, {Devost}, \&
  {Heckman}}]{1999ApJS..123....3L}
{Leitherer}, C., {Schaerer}, D., {Goldader}, J.~D., {et~al.} 1999, \apjs, 123,
  3

\bibitem[{{Mart{\'{\i}}n-Navarro} {et~al.}(2012){Mart{\'{\i}}n-Navarro},
  {Bakos}, {Trujillo}, {Knapen}, {Athanassoula}, {Bosma}, {Comer{\'o}n},
  {Elmegreen}, {Erroz-Ferrer}, {Gadotti}, {Gil de Paz}, {Hinz}, {Ho},
  {Holwerda}, {Kim}, {Laine}, {Laurikainen}, {Men{\'e}ndez-Delmestre},
  {Mizusawa}, {Mu{\~n}oz-Mateos}, {Regan}, {Salo}, {Seibert}, \&
  {Sheth}}]{2012MNRAS.427.1102M}
{Mart{\'{\i}}n-Navarro}, I., {Bakos}, J., {Trujillo}, I., {et~al.} 2012,
  \mnras, 427, 1102

\bibitem[{{Mart{\'{\i}}nez-Serrano} {et~al.}(2009){Mart{\'{\i}}nez-Serrano},
  {Serna}, {Dom{\'e}nech-Moral}, \&
  {Dom{\'{\i}}nguez-Tenreiro}}]{2009ApJ...705L.133M}
{Mart{\'{\i}}nez-Serrano}, F.~J., {Serna}, A., {Dom{\'e}nech-Moral}, M., \&
  {Dom{\'{\i}}nguez-Tenreiro}, R. 2009, \apjl, 705, L133

\bibitem[{{Minchev} \& {Famaey}(2010)}]{2010ApJ...722..112M}
{Minchev}, I. \& {Famaey}, B. 2010, \apj, 722, 112

\bibitem[{{Minchev} {et~al.}(2012){Minchev}, {Famaey}, {Quillen}, {Di Matteo},
  {Combes}, {Vlaji{\'c}}, {Erwin}, \& {Bland-Hawthorn}}]{2012A&A...548A.126M}
{Minchev}, I., {Famaey}, B., {Quillen}, A.~C., {et~al.} 2012, \aap, 548, A126

\bibitem[{{Patterson}(1940)}]{1940BHarO.914....9P}
{Patterson}, F.~S. 1940, Harvard College Observatory Bulletin, 914, 9

\bibitem[{{P{\'e}rez}(2004)}]{2004A&A...427L..17P}
{P{\'e}rez}, I. 2004, \aap, 427, L17

\bibitem[{{Pilkington} {et~al.}(2012){Pilkington}, {Few}, {Gibson}, {Calura},
  {Michel-Dansac}, {Thacker}, {Moll{\'a}}, {Matteucci}, {Rahimi}, {Kawata},
  {Kobayashi}, {Brook}, {Stinson}, {Couchman}, {Bailin}, \&
  {Wadsley}}]{2012A&A...540A..56P}
{Pilkington}, K., {Few}, C.~G., {Gibson}, B.~K., {et~al.} 2012, \aap, 540, A56

\bibitem[{{Pohlen} {et~al.}(2002){Pohlen}, {Dettmar}, {L{\"u}tticke}, \&
  {Aronica}}]{2002A&A...392..807P}
{Pohlen}, M., {Dettmar}, R.-J., {L{\"u}tticke}, R., \& {Aronica}, G. 2002,
  \aap, 392, 807

\bibitem[{{Pohlen} \& {Trujillo}(2006)}]{2006A&A...454..759P}
{Pohlen}, M. \& {Trujillo}, I. 2006, \aap, 454, 759

\bibitem[{{Quillen} {et~al.}(2009){Quillen}, {Minchev}, {Bland-Hawthorn}, \&
  {Haywood}}]{2009MNRAS.397.1599Q}
{Quillen}, A.~C., {Minchev}, I., {Bland-Hawthorn}, J., \& {Haywood}, M. 2009,
  \mnras, 397, 1599

\bibitem[{{Roediger} {et~al.}(2012){Roediger}, {Courteau},
  {S{\'a}nchez-Bl{\'a}zquez}, \& {McDonald}}]{2012ApJ...758...41R}
{Roediger}, J.~C., {Courteau}, S., {S{\'a}nchez-Bl{\'a}zquez}, P., \&
  {McDonald}, M. 2012, \apj, 758, 41

\bibitem[{{Ro{\v s}kar} {et~al.}(2008{\natexlab{a}}){Ro{\v s}kar},
  {Debattista}, {Quinn}, {Stinson}, \& {Wadsley}}]{2008ApJ...684L..79R}
{Ro{\v s}kar}, R., {Debattista}, V.~P., {Quinn}, T.~R., {Stinson}, G.~S., \&
  {Wadsley}, J. 2008{\natexlab{a}}, \apjl, 684, L79

\bibitem[{{Ro{\v s}kar} {et~al.}(2008{\natexlab{b}}){Ro{\v s}kar},
  {Debattista}, {Stinson}, {Quinn}, {Kaufmann}, \&
  {Wadsley}}]{2008ApJ...675L..65R}
{Ro{\v s}kar}, R., {Debattista}, V.~P., {Stinson}, G.~S., {et~al.}
  2008{\natexlab{b}}, \apjl, 675, L65

\bibitem[{{Ruiz-Lara} {et~al.}(2016{\natexlab{a}}){Ruiz-Lara}, {Few}, {Gibson},
  {P{\'e}rez}, {Florido}, {Minchev}, \&
  {S{\'a}nchez-Bl{\'a}zquez}}]{2016A&A...586A.112R}
{Ruiz-Lara}, T., {Few}, C.~G., {Gibson}, B.~K., {et~al.} 2016{\natexlab{a}},
  \aap, 586, A112

\bibitem[{{Ruiz-Lara} {et~al.}(2016{\natexlab{b}}){Ruiz-Lara}, {P{\'e}rez},
  {Florido}, {S{\'a}nchez-Bl{\'a}zquez}, {M{\'e}ndez-Abreu}, {Lyubenova},
  {Falc{\'o}n-Barroso}, {S{\'a}nchez-Menguiano}, {S{\'a}nchez}, {Galbany},
  {Garc{\'{\i}}a-Benito}, {Gonz{\'a}lez Delgado}, {Husemann}, {Kehrig},
  {L{\'o}pez-S{\'a}nchez}, {Marino}, {Mast}, {Papaderos}, {van de Ven},
  {Walcher}, {Zibetti}, \& {CALIFA Team}}]{2016MNRAS.456L..35R}
{Ruiz-Lara}, T., {P{\'e}rez}, I., {Florido}, E., {et~al.} 2016{\natexlab{b}},
  \mnras, 456, L35

\bibitem[{{Ruiz-Lara} {et~al.}(2017){Ruiz-Lara}, {P{\'e}rez}, {Florido},
  {S{\'a}nchez-Bl{\'a}zquez}, {M{\'e}ndez-Abreu}, {S{\'a}nchez-Menguiano},
  {S{\'a}nchez}, {Lyubenova}, {Falc{\'o}n-Barroso}, {van de Ven}, {Marino}, {de
  Lorenzo-C{\'a}ceres}, {Catal{\'a}n-Torrecilla}, {Costantin},
  {Bland-Hawthorn}, {Galbany}, {Garc{\'{\i}}a-Benito}, {Husemann}, {Kehrig},
  {M{\'a}rquez}, {Mast}, {Walcher}, {Zibetti}, {Ziegler}, \& {Califa
  Team}}]{2017A&A...604A...4R}
{Ruiz-Lara}, T., {P{\'e}rez}, I., {Florido}, E., {et~al.} 2017, \aap, 604, A4

\bibitem[{{S{\'a}nchez} {et~al.}(2012){S{\'a}nchez}, {Kennicutt}, {Gil de Paz},
  {van de Ven}, {V{\'{\i}}lchez}, {Wisotzki}, {Walcher}, {Mast}, {Aguerri},
  {Albiol-P{\'e}rez}, {Alonso-Herrero}, {Alves}, {Bakos}, {Bart{\'a}kov{\'a}},
  {Bland-Hawthorn}, {Boselli}, {Bomans}, {Castillo-Morales}, {Cortijo-Ferrero},
  {de Lorenzo-C{\'a}ceres}, {Del Olmo}, {Dettmar}, {D{\'{\i}}az}, {Ellis},
  {Falc{\'o}n-Barroso}, {Flores}, {Gallazzi}, {Garc{\'{\i}}a-Lorenzo},
  {Gonz{\'a}lez Delgado}, {Gruel}, {Haines}, {Hao}, {Husemann},
  {Igl{\'e}sias-P{\'a}ramo}, {Jahnke}, {Johnson}, {Jungwiert}, {Kalinova},
  {Kehrig}, {Kupko}, {L{\'o}pez-S{\'a}nchez}, {Lyubenova}, {Marino},
  {M{\'a}rmol-Queralt{\'o}}, {M{\'a}rquez}, {Masegosa}, {Meidt},
  {Mendez-Abreu}, {Monreal-Ibero}, {Montijo}, {Mour{\~a}o}, {Palacios-Navarro},
  {Papaderos}, {Pasquali}, {Peletier}, {P{\'e}rez}, {P{\'e}rez}, {Quirrenbach},
  {Rela{\~n}o}, {Rosales-Ortega}, {Roth}, {Ruiz-Lara},
  {S{\'a}nchez-Bl{\'a}zquez}, {Sengupta}, {Singh}, {Stanishev}, {Trager},
  {Vazdekis}, {Viironen}, {Wild}, {Zibetti}, \&
  {Ziegler}}]{2012A&A...538A...8S}
{S{\'a}nchez}, S.~F., {Kennicutt}, R.~C., {Gil de Paz}, A., {et~al.} 2012,
  \aap, 538, A8

\bibitem[{{S{\'a}nchez-Bl{\'a}zquez} {et~al.}(2009){S{\'a}nchez-Bl{\'a}zquez},
  {Courty}, {Gibson}, \& {Brook}}]{2009MNRAS.398..591S}
{S{\'a}nchez-Bl{\'a}zquez}, P., {Courty}, S., {Gibson}, B.~K., \& {Brook},
  C.~B. 2009, \mnras, 398, 591

\bibitem[{{S{\'a}nchez-Menguiano} {et~al.}(2016){S{\'a}nchez-Menguiano},
  {S{\'a}nchez}, {Kawata}, {Chemin}, {P{\'e}rez}, {Ruiz-Lara},
  {S{\'a}nchez-Bl{\'a}zquez}, {Galbany}, {Anderson}, {Grand}, {Minchev}, \&
  {G{\'o}mez}}]{2016ApJ...830L..40S}
{S{\'a}nchez-Menguiano}, L., {S{\'a}nchez}, S.~F., {Kawata}, D., {et~al.} 2016,
  \apjl, 830, L40

\bibitem[{{Sellwood} \& {Binney}(2002)}]{2002MNRAS.336..785S}
{Sellwood}, J.~A. \& {Binney}, J.~J. 2002, \mnras, 336, 785

\bibitem[{{Steinmetz} \& {Navarro}(2002)}]{2002NewA....7..155S}
{Steinmetz}, M. \& {Navarro}, J.~F. 2002, \na, 7, 155

\bibitem[{{Teyssier}(2002)}]{2002A&A...385..337T}
{Teyssier}, R. 2002, \aap, 385, 337

\bibitem[{{Thompson} {et~al.}(2017){Thompson}, {Few}, {Bergemann}, {Gibson},
  {MacFarlane}, {Serenelli}, {Gilmore}, {Randich}, {Vallenari}, {Alfaro},
  {Bensby}, {Francois}, {Korn}, {Bayo}, {Carraro}, {Casey}, {Costado},
  {Donati}, {Franciosini}, {Frasca}, {Hourihane}, {Jofre}, {Hill}, {Heiter},
  {Koposov}, {Lanzafame}, {Lardo}, {de Laverny}, {Lewis}, {Magrini}, {Marconi},
  {Masseron}, {Monaco}, {Morbidelli}, {Pancino}, {Prisinzano}, {Recio-Blanco},
  {Sacco}, {Sousa}, {Tautvaisiene}, {Worley}, \&
  {Zaggia}}]{2017arXiv170901523T}
{Thompson}, B.~B., {Few}, C.~G., {Bergemann}, M., {et~al.} 2017, ArXiv e-prints
  [\eprint[arXiv]{1709.01523}]

\bibitem[{{Trujillo} \& {Pohlen}(2005)}]{2005ApJ...630L..17T}
{Trujillo}, I. \& {Pohlen}, M. 2005, \apjl, 630, L17

\bibitem[{{van der Kruit}(1987)}]{1987A&A...173...59V}
{van der Kruit}, P.~C. 1987, \aap, 173, 59

\bibitem[{{White} \& {Rees}(1978)}]{1978MNRAS.183..341W}
{White}, S.~D.~M. \& {Rees}, M.~J. 1978, \mnras, 183, 341

\bibitem[{{Yoachim} {et~al.}(2012){Yoachim}, {Ro{\v s}kar}, \&
  {Debattista}}]{2012ApJ...752...97Y}
{Yoachim}, P., {Ro{\v s}kar}, R., \& {Debattista}, V.~P. 2012, \apj, 752, 97

\bibitem[{{Younger} {et~al.}(2007){Younger}, {Cox}, {Seth}, \&
  {Hernquist}}]{2007ApJ...670..269Y}
{Younger}, J.~D., {Cox}, T.~J., {Seth}, A.~C., \& {Hernquist}, L. 2007, \apj,
  670, 269

\end{thebibliography}

 
\end{document}